\documentclass{PoS}
\usepackage{enumerate}
\usepackage{amssymb}
\usepackage{amsmath}
\title{Two-loop-induced neutrino masses: A model-independent
  perspective}

\ShortTitle{Two-loop-induced neutrino masses}

\author{\speaker{D. Aristizabal Sierra}
  \thanks{Talk presented at the 18th International Conference From the
    Planck Scale to the Electroweak Scale}
  \\
  IFPA, Dep. AGO, Universit\'e de Li\`ege, Bat B5, Sart Tilman B-4000
  Li\`ege 1, Belgium
  \\
  E-mail: \email{daristizabal@ulg.ac.be}}

\abstract{We discuss Majorana neutrino mass generation mechanisms at
  the two-loop order. After briefly reviewing the systematic
  classification of one-loop realizations, we then focus on a general
  two-loop classification scheme which provides a model-independent
  catalog for neutrino mass models at the two-loop order.}

\FullConference{18th International Conference From the Planck Scale 
  to the Electroweak Scale \\
  25-29 May 2015\\
  Ioannina, Greece }

\begin{document}

\section{Introduction}
\label{sec:intro}
Results from neutrino oscillation experiments have demonstrated that
neutrinos are massive and that the lepton mixing pattern follows a
non-trivial structure
\cite{Fukuda:1998mi,Ahmad:2002jz,Eguchi:2002dm}. These results,
combined with the cosmic baryon asymmetry, and cosmological and
astrophysical data supporting the hypothesis of dark matter, are of
for today the most solid evidence for beyond-the-standard-model (BSM)
physics.

Values for neutrino mass square differences are derived from
oscillation data \cite{Forero:2014bxa}, while upper limits on the
absolute neutrino mass are given by neutrinoless double-beta decay
\cite{Auger:2012ar}, kinematic endpoint measurements of the $\beta$
decay spectrum \cite{Kraus:2004zw} and cosmology
\cite{Ade:2015xua}. They all show that neutrino masses are tiny, with
values way below the masses of the other standard model (SM) fermions,
$\mathcal{O}\sim 0.1\,$~eV. Neutrinos, being electrically neutral, are
the only SM particles that can have Majorana nature. Testing whether
this is the case, requires observing lepton-number-violating (LNV)
processes, of which neutrinoless double-beta decay is probably the
most promising signature (see
e.g. \cite{Rodejohann:2012xd,Deppisch:2012nb} for a details
discussion). If neutrinos are Majorana particles, their masses will be
generated differently compared to the other SM fermions. This, indeed,
might be the fundamental reason behind the smallness of neutrino
masses.

At the effective level, Majorana neutrino masses can be generated by
the dimension five effective operator $\mathcal{O}_5\sim
\ell\,\ell\,H\,H$, the Weinberg operator \cite{Weinberg:1980bf}. Such
a description suffices to account for neutrino oscillation data,
however understanding the origin of neutrino masses requires going
beyond this effective description and calls for unclosing the nature
of the UV completion responsible for this operator.

It is worth pointing out that although a particular realization for
$\mathcal{O}_5$ (or any higher-order LNV operator) fixes the mechanism
for neutrino mass generation, by itself does not provide any
information about the origin of the neutrino mixing pattern. The
conventional procedure to tackle this problem consist on embedding the
model into a larger framework which involves a flavor symmetry. In
that context, neutrino mixing is understood as a consequence of flavor
symmetry breaking (see refs. \cite{King:2015aea,Morisi:2012fg} for
further details). Another interesting approach is that in which the
mixing, assumed to be in first approximation of TBM form, results from
different mechanisms contributing to the neutrino mass matrix. The
main idea is that while one of the mechanisms accounts for the TBM
structure, the other one is only responsible for the deviations (see
refs. \cite{Sierra:2013ypa}). It is conceivable as well that
deviations from the TBM form, and even the origin of CP violation in
the lepton sector, arise from deviations from minimality. This can be
readily seen e.g. in the type-I seesaw endowed with an $A_4$ symmetry
\cite{Altarelli:2005yx}. Leading-order terms lead to an exact TBM
pattern, additional right-handed neutrinos (and not higher-order
effective operators) could---in principle---lead to the experimentally
required deviations.
\section{Classification schemes}
\label{sec:systematics}
Majorana neutrino mass models can be regarded as ``incarnations'' of
either the Weinberg operator or another higher-order and LNV operator
\cite{Bonnet:2009ej}. The number of varieties is extremely large, and
covers UV completions which due to their particle content (or the
presence of extra symmetries) generate a neutrino mass matrix either
at the tree level or at higher order. For models associated with
$\mathcal{O}_5$, possibilities ranging from tree level up to
three-loop level have been considered. Tree level realizations
correspond to the standard seesaw models, type-I, II and III. Beyond
the tree level, the number of models that have been discussed in the
literature is huge (see references in
\cite{Bonnet:2012kz,Sierra:2014rxa} for an almost ``complete''
list). Examples are the Zee model \cite{Zee:1980ai}, models with
colored scalars \cite{AristizabalSierra:2007nf,FileviezPerez:2009ud}
and radiative seesaw models \cite{Ma:2006km} \footnote{``Hybrid''
  neutrino mass models where neutrino masses are generated by tree and
  one-loop contributions can be considered in the type-I seesaw
  itself, see refs.
  \cite{Grimus:1989pu,Pilaftsis:1991ug,AristizabalSierra:2011mn}. Thus,
  they could be included in the one-loop realizations too.}. Detailed
phenomenological analyses of these realizations have been done, in
particular for the Zee and the radiative seesaw models, see
e.g. \cite{AristizabalSierra:2006ri,Sierra:2008wj}. For two-loop
models one could mention e.g. the Cheng-Li-Babu-Zee model
\cite{Cheng:1980qt,Zee:1985id,Babu:1988ki}, which has been the subject
of extensive phenomenological studies (see
e.g. \cite{AristizabalSierra:2006gb,Nebot:2007bc,Herrero-Garcia:2014hfa,
  Schmidt:2014zoa}). Two-loop snail models, recently pointed out in
\cite{Farzan:2014aca}, are as well examples of two-loop neutrino mass
generation.  At the three-loop order, as an example one could mention
the ``cocktail'' model \cite{Gustafsson:2012vj}.

Construction of models based on specific Lagrangians leads to a
certain degree of arbitrariness. This can---in principle---be avoided
if one relies on the following observation: There is a vast class of
models which are just UV completions of $\mathcal{O}_5$. Thus, rather
than writing particular Lagrangians one could think of starting with a
``unique'' object and systematically study all its realizations at
different orders. Relying on a diagrammatic approach, such a program
has been pursued up to the two-loop order
\cite{Ma:1998dn,Bonnet:2012kz,Sierra:2014rxa}. Other attempts aiming
at systematic classifications have been presented in
\cite{Babu:2001ex,Choi:2002bb,deGouvea:2007xp,delAguila:2012nu,Angel:2012ug,Farzan:2012ev}.

The diagrammatic approach follows a simple strategy which can be
summarized as follows. As soon as the order at which the analysis is
going to be done is fixed: $(i)$ find the topologies that can
potentially lead to ``genuine'' diagrams, $(ii)$ from those topologies
draw the different diagrams paying special attention to those that can
``genuinely'' determine the mass matrix, $(iii)$ fix the EW quantum
numbers of the BSM fields, $(iv)$ calculate the loop integrals. The
list of models resulting from such classification provide a catalog of
``genuine'' neutrino mass models. With genuine here we refer to
diagrams for which the absence of leading-order diagrams can be
guaranteed.  Only in those cases it can be entirely assured that the
neutrino mass matrix originates at the same order that the
corresponding diagram.
\section{Brief review of systematic classification of one-loop
  neutrino mass models}
\label{sec:one-loop}
\begin{figure}[t]
  \centering
  \includegraphics[scale=0.65]{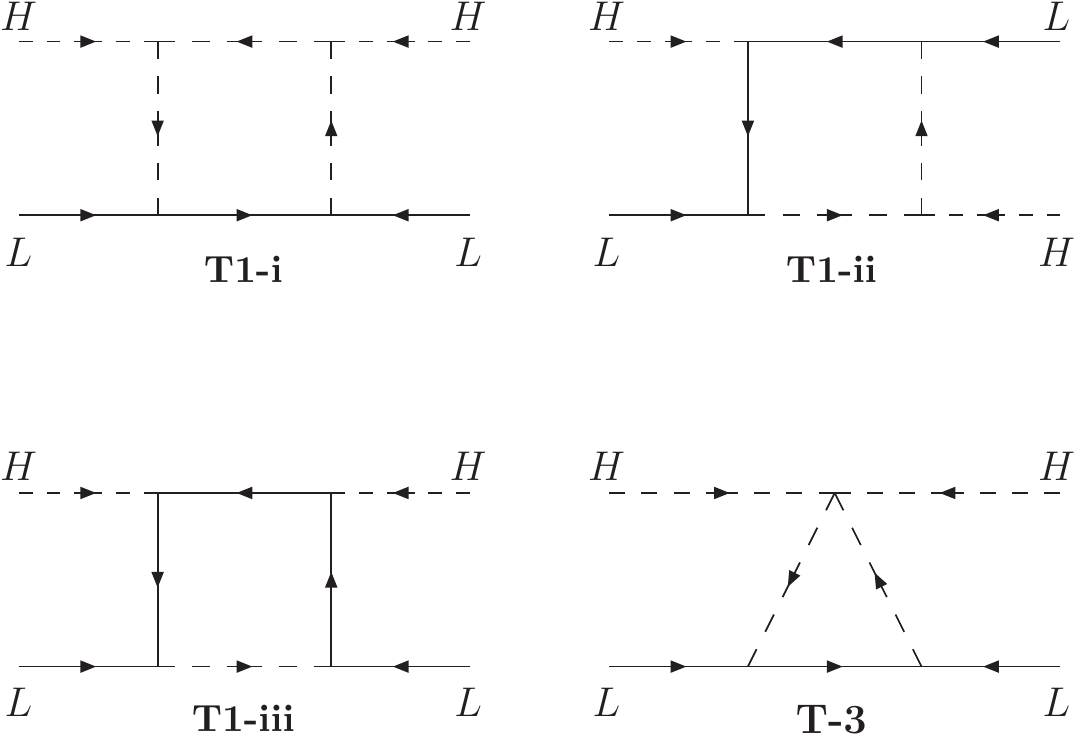}
  \caption{One-loop genuine diagrams according to
    ref. \cite{Bonnet:2012kz}.}
  \label{fig:one-loop}
\end{figure}

A diagrammatic-based systematic classification of $\mathcal{O}_5$ at
the one-loop order has been presented in \cite{Bonnet:2012kz}. In
there it was shown that out of the six possible topologies only the
one particle-irreducibles (1PI) lead to genuine diagrams, shown in
fig. \ref{fig:one-loop}. It can be noted that with appropriate quantum
number assignments, diagrams T1-ii will lead to the Zee model, while
T-3 to the scotogenic model.

A one particle-reducible topology of interest is found in this
case. It leads to a set of non-genuine but finite diagrams that can be
understood as originating from the type-I, II or III tree level
diagrams (depending on the diagram) where one of the vertices
(couplings) is one-loop induced. In that sense they can be regarded as
``effective'' one-loop diagrams. They are of relevance since in those
models a small parameter entering in the neutrino mass matrix is
justified by its loop-induced nature. EW quantum numbers for the
relevant diagrams \footnote{In the one-loop case the hypercharge of
  the BSM fields is fixed up to an arbitrary constant, $\alpha$.} as
well as explicit expressions for the different one-loop integrals can
be found in \cite{Bonnet:2012kz}.
\section{Two-loop classification scheme and catalog}
\label{sec:two-loop}
\begin{figure}[t]
  \centering
  \includegraphics[scale=0.75]{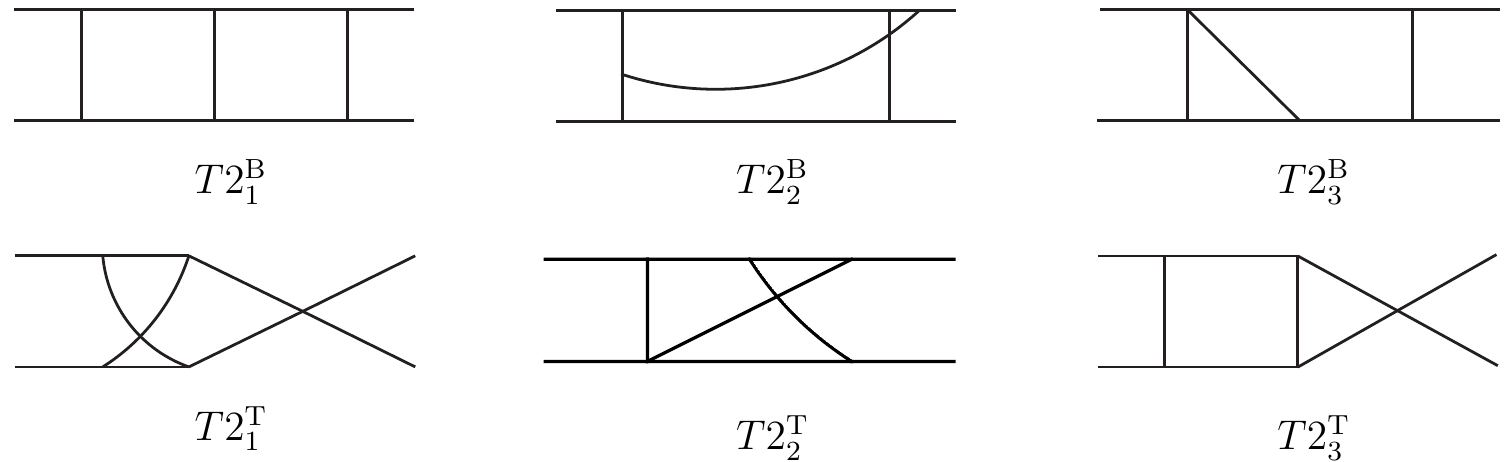}
  \caption{Two-loop topologies that potentially lead to genuine
    diagrams \cite{Sierra:2014rxa}.}
  \label{fig:relevant-topo}
\end{figure}

The two-loop diagrammatic analysis has been presented in
ref. \cite{Sierra:2014rxa}. In that case the number of topologies is
of course larger, something that in turn implies a larger number of
diagrams and therefore models. Out of the 29 topologies, only six are
found to potentially lead to genuine diagrams, as shown in
fig. \ref{fig:relevant-topo}.

Non-genuine but finite diagrams are found as well. Some of them arise
from the topologies in fig. \ref{fig:relevant-topo}, but some others
from topologies which we here are not listing (for the full list see
ref. \cite{Sierra:2014rxa}). As in the one-loop case, they are of
interest as well since enable understanding the smallness of small
couplings entering in the mass matrix.

To guarantee that the diagrams one gets from the topologies in
fig. \ref{fig:relevant-topo} are truly genuine, additional
model-construction rules have to be used, namely:
\begin{enumerate}[(I)]
\item The resulting particle content should not include 
hypercharge zero fermion EW singlets and triplets or hypercharge two scalar
EW triplets.
\item The resulting particle content should not contain either
  hypercharge zero scalar $SU(2)$ singlets or triplets.
\item Depending on the diagram, BSM scalars should not carry the same
  quantum numbers than the SM Higgs. In other cases they can, but
  subject to conditions on the internal Yukawa vertices (see
  \cite{Sierra:2014rxa} for a detailed discussion).
\item For the quartic couplings in diagram T-3 in
  fig. \ref{fig:one-loop} the following field assignments:
  $S_{1,2}=S_D$, $S_1=S_S$ and $S_2=S_T$, $S_1=S_T$ and $S_2=S_T$
  (where with $T,D,S$ we refer to EW triplets, doublets and singlets),
  are possible only if $|Y(S_1)-Y(S_2)|\neq Y_H$, with $Y_H$ referring
  to the Higgs hypercharge (this rule has an exception, see
  ref. \cite{Sierra:2014rxa}).
\end{enumerate}

\begin{figure}
  \centering
  \includegraphics[scale=0.7]{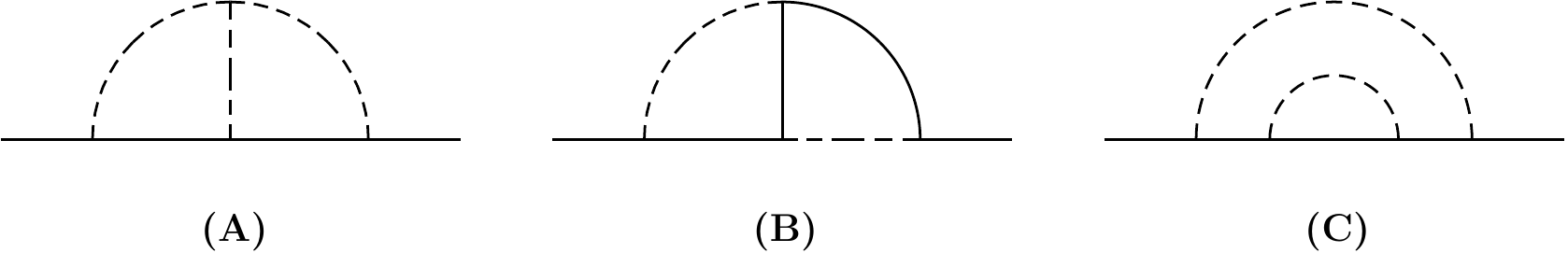}
  \caption{Different classes of genuine two-loop diagrams. Depending
    on how the scalar external legs are attached different ``models''
    are obtained.}
  \label{fig:diagrams-gen}
\end{figure}
When complemented with these rules, the topologies in
fig. \ref{fig:relevant-topo} lead to three, and only three, different
categories of genuine diagrams, as shown in
fig. \ref{fig:diagrams-gen}. Depending on how the attachment of the
external Higgs lines is done, different diagrams arise. For diagrams
in category (A) we have found 10 different possibilities, for (B) six
and for (C) four.

With the diagrams at hand, models for two-loop neutrino mass
generation can be written provided quantum numbers are specified. For
genuine diagrams, this has been given in ref. \cite{Sierra:2014rxa}
for EW representations up to triplets. Note that in these models
hypercharge for the BSM fields is determined up to two arbitrary
constants, $\alpha$ and $\beta$.

The two-loop integrals one can found depend on the chiral structure of
internal vertices. However, in the most general case of internal
vector-like couplings, the problem involves just four different types
of integrals. They were studied long time ago in
\cite{vanderBij:1983bw}, and were adapted to the different possible
models in \cite{Sierra:2014rxa}.
\subsection{Using these results to construct two-loop models}
\label{sec:using-this-results}
\begin{figure}[t]
  \centering
  \includegraphics[scale=0.65]{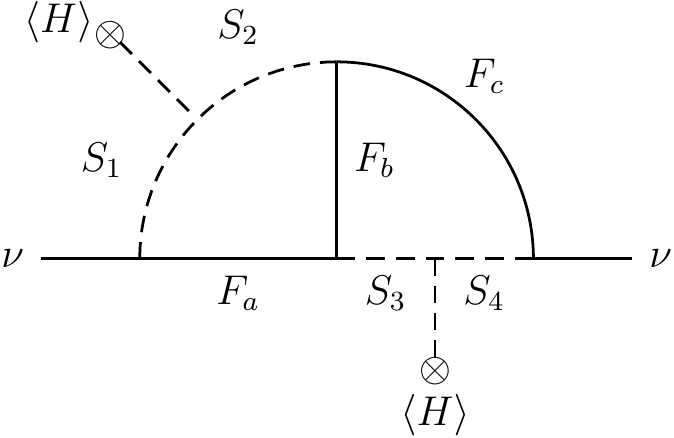}
  \caption{Genuine two-loop model example. The diagram falls into
    category (B).}
  \label{fig:genuine-exam}
\end{figure}

In what follows we illustrate how the two-loop model-building catalog
of ref. \cite{Sierra:2014rxa} can be used by means of an example. We
believe this is more ``illuminating'' than just displaying tables with
EW quantum numbers. The example is based on a diagram falling in
category (B), as shown in fig. \ref{fig:genuine-exam}. The quantum
numbers for the BSM fields were taken from the tab. 3 in
\cite{Sierra:2014rxa}, and are displayed in
tab. \ref{tab:QN-example}. They correspond to the hypercharge
constants choice $\alpha=2$ and $\beta=-3$.

The diagram in fig.~\ref{fig:genuine-exam} follows from the Lagrangian
\begin{equation}
  \label{eq:lag-ptbm-3}
  {\cal L}_\text{int} =
  Y_{ia}\,\left(\overline{L^c_i} P_L S_1 \right)\cdot F^c_a
  +
  Y_{cj}\,\left(\overline{F_c} P_L L_j\right) \cdot S_4
  +
  h_{ab}\,\overline{F^c_a} \cdot \left(F^c_b S_3^\dagger\right)
  +
  h_{bc}\,\left(\overline{F^c_b} F_c\right) \cdot S_2^\dagger
  +
  \mbox{H.c.}\ ,
\end{equation}
and the scalar potential:
\begin{equation}
  \label{eq:scalar-pot-ptbm-3}
  V \supset
  \mu_{34}\,S_4^\dagger \cdot \left(S_3 H \right)
  +
  \mu_{12}\,S_2\cdot \left(S_1^\dagger H\right)
  +
  \mbox{H.c.}
  +
  \sum_{x=1}^4 m_{S_x}^2 \left|S_x\right|^2\, ,
\end{equation}
where the parenthesis refer to $SU(2)$ contractions. The fermions,
being vector-like, allow for the following terms
\begin{equation}
  \label{eq:vectorlike-mass-terms-ptbm-3-exam}
  {\cal L}_M = \sum_{A=a,b,c} m_{F_A}\overline{F_A}F_A\ .
\end{equation}

\begin{table}[t]
  \renewcommand{\arraystretch}{1.2}
  \setlength{\tabcolsep}{4pt}
  \centering
  \begin{tabular}{|c||c|c|c|c|c|c|c|}
    \hline
    \multicolumn{8}{|c|}{\bf Category (B) model example}
    \\\hline
    {\sc Fields}& $F_a$ & $F_b$ & $F_c$ & $S_1$ & $S_2$ & $S_3$ & $S_4$
    \\
    \hline
    $SU(2)_L$ & 1 & 2 & 2 & 2 & 1 & 2 & 1
    \\
    $U(1)_Y$ & 1 & 5 & $-4$ & 2 & 1 & $-4$ & $-3$ 
    \\\hline
  \end{tabular}
  \caption{Quantum numbers for the two-loop example.}
  \label{tab:QN-example}
\end{table}
\noindent
Coupling $\mu_{34}$ in (\ref{eq:scalar-pot-ptbm-3}) implies mixing
between the $Q=3/2$ scalars, while $\mu_{12}$ between the
$Q=1/2$ states. The mass matrices can then be written as
\begin{equation}
  \label{eq:scalar-mass-matrix-ptbm-3-EX}
  M_{S^{Q=3/2}}^2=
  \begin{pmatrix}
    m_{S_3}^2   & \mu_{34}v\\
    \mu_{34}v & m_{S_4}^2
  \end{pmatrix}\ ,
  \qquad
  M_{S^{Q=1/2}}^2=
  \begin{pmatrix}
    m_{S_1}^2   & \mu_{12}v\\
    \mu_{12}v   & m_{S_2}^2
  \end{pmatrix}\ .
\end{equation}
These matrices can be diagonalized with $2\times 2$ rotation matrices,
namely
\begin{equation}
  \label{eq:rot-mat-ptbm-3}
  R_{Q}=
  \begin{pmatrix}
    \cos\theta_Q & \sin \theta_Q\\
    -\sin\theta_Q & \cos\theta_Q
  \end{pmatrix}\, ,
\end{equation}
with the rotation angles written as
\begin{equation}
  \label{eq:roatation-angles-ptbm-3}
  \tan 2\theta_{Q=3/2}=\frac{2\mu_{34}v}{m_{S_3}^2 - m_{S_4}^2}\ ,
  \qquad
  \tan 2\theta_{Q=1/2}=\frac{2\mu_{12}v}{m_{S_1}^2 - m_{S_2}^2}\, ,
\end{equation}
where $v\equiv \langle H\rangle$.  The full neutrino mass matrix is
calculated after rotation of the interactions in (\ref{eq:lag-ptbm-3})
and (\ref{eq:scalar-pot-ptbm-3}) to the scalar mass eigenstate basis.

\begin{figure}[t]
  \centering
  \includegraphics[scale=0.55]{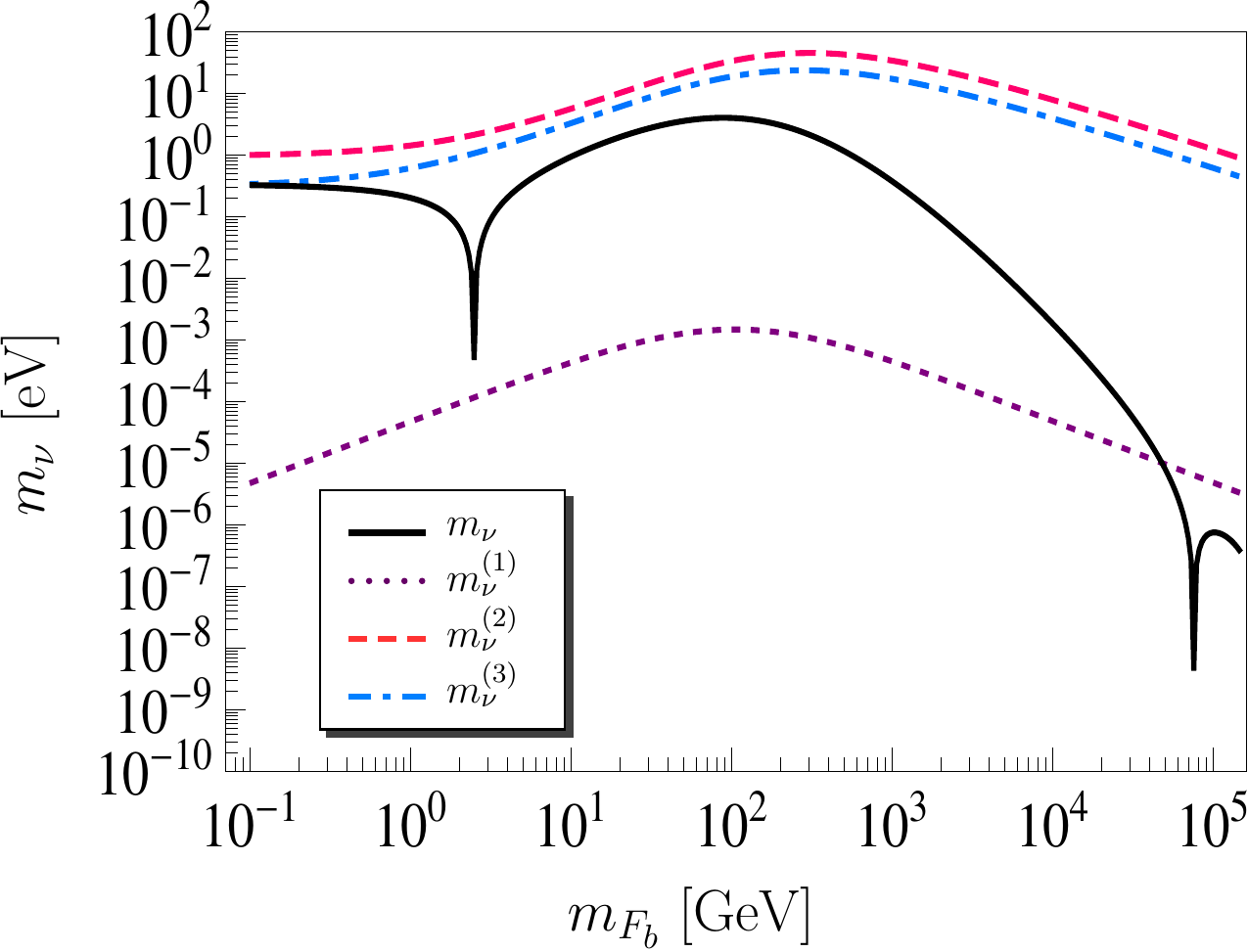}
  \hfill
  \includegraphics[scale=0.55]{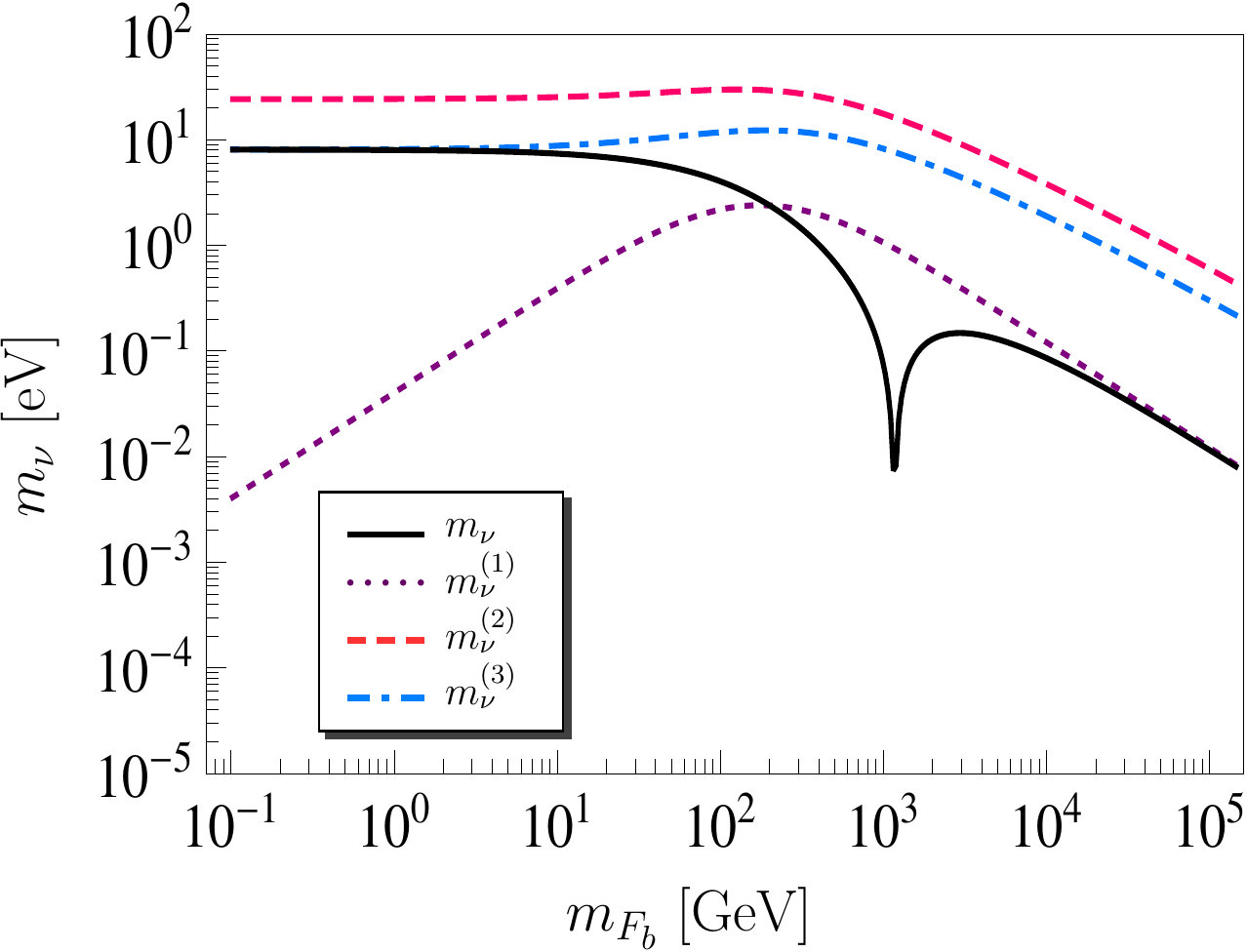}
  \caption{Behavior of the different two-loop integrals at the
    neutrino mass matrix level.  The solid black curve shows the total
    $m_\nu$, the other curves the individual contributions
    $m_\nu^{(i)}$, determined by the functions
    $F^{(i)}_{ab,\alpha\beta,b}$ (see the text).  These plots aim only
    at showing the behavior of the different integrals. They do not
    represent any phenomenological analysis, both the overall neutrino
    scale and the values of $m_{F_b}$ have to be understood in that
    way.}
  \label{fig:integral-plot}
\end{figure}

The chiral structure of the external vertices for the diagram in
fig. \ref{fig:genuine-exam} is fixed by the SM to be
$P_L=(1-\gamma_5)/2$. Since the BSM fermions are vector-like, the
internal vertices of the diagram are not chiral. Bearing that in mind,
one can then write the neutrino mass matrix as follows:
\begin{align}
  \label{eq:neutrino-mm-ptbm-3}
  {\cal M}_\nu=\frac{1}{4(16\pi^2)^2}\left(Y_{ia}Y_{cj} + Y_{ja}Y_{ci}\right)
  h_{ab}h_{bc}\sin 2\theta_{Q=3/2}\sin 2\theta_{Q=1/2}
  \sum_{A=1}^4 \sum_{\alpha,\beta}(-1)^\alpha(-1)^\beta F_{ac,\alpha\beta,b}^{(A)}\ ,
\end{align}
where the dimensionful functions $F_{ac,\alpha\beta,b}^{(A)}$ are
determined by four types of two-loop integrals:
\begin{align}
  \label{eq:F1}
  F^{(1)}_{ab,\alpha\beta,b}&=\frac{m_{F_a}m_{F_c}}{m_{F_b}}\,
  \times\pi^{-4}\,\hat {\cal I}_{ac,\alpha\beta}\ ,
  \\
  \label{eq:F2}
  F^{(2)}_{ab,\alpha\beta,b}&=\left(m_{F_a} + m_{F_b} + m_{F_c}\right)\,
  \times\pi^{-4}\,\hat {\cal I}_{ac,\alpha\beta}^{[(k+q)^2]}\ ,
  \\
  \label{eq:F3}
  F^{(3)}_{ac,\alpha\beta,b}&=-(m_{F_a} + m_{F_b})\,
  \times\pi^{-4}\,\hat {\cal I}_{ac,\alpha\beta}^{(k^2)}\ ,
  \\
  \label{eq:F4}
  F^{(4)}_{ac,\alpha\beta,b}&=-(m_{F_b} + m_{F_c})\,
  \times\pi^{-4}\,\hat {\cal I}_{ac,\alpha\beta}^{(q^2)}\, .
\end{align}
Rather than given explicit expressions for the integrals
$\hat{\mathcal{I}}$, for which the reader is referred to
\cite{Sierra:2014rxa}, we display their effect in the neutrino mass
matrix as a function of the internal fermion mass $F_b$ in
fig. \ref{fig:integral-plot}.  Note that in this result we have not
consider any flavor structure and we have fixed all the Yukawa
couplings to 1. In that sense, these results are just representative
of the typical overall neutrino mass and have nothing to do with any
possible prediction of this model.

The plots were done for scalar mass parameters $m_{S_1}^2 = 100^2$
GeV$^2$ and $m_{S_2}^2= m_{S_1}^2 + \Delta m^2$ (with $\Delta m^2
=\mu\,v$) and fixed $\Delta m^2 =24.6$ GeV$^2$ and two different
values of $m_F=m_{F_a}=m_{F_c}$: to the left 1 GeV and to the right
100 GeV. The black curve shows the full $m_{\nu}$, the other curves
show, instead, the different contributions $m^{(i)}_\nu$, $i=1,2,3$
individually (determined by the functions
in~(\ref{eq:F1})-(\ref{eq:F3}) and the common global factor in
(\ref{eq:neutrino-mm-ptbm-3})).  We have found that $m^{(4)}_\nu$ is
numerically equal to $m^{(3)}_\nu$, while $m^{(2)}_\nu < 0$ and we
plot the absolute value. Usually the contribution from
$m^{(2)}_\nu-m^{(3)}_\nu$ dominates the neutrino mass for small and
moderate values of $m_{F_b}$, but at large values of $m_{F_b}$,
$m_\nu^{(2)}$ and $m_\nu^{(3)}+m_\nu^{(4)}$ tend to cancel each other,
such that the only remaining contribution comes from $m_\nu^{(1)}$. In
the plots there are some points for $m_{F_b}$, for which the different
contributions can actually exactly cancel each other. Note also, that
for $m_{F_b} \to \infty$, $m_{\nu}$ goes to zero, as
expected. Obviously, as these plots demonstrate, neutrino masses of
the correct order of magnitude can be achieved for a wide range of
input parameters.

\section{Conclusions}
\label{sec:concl}
After briefly reviewing the systematic classification of the Weinberg
operator at the one-loop order, we discussed a general classification
for the two-loop case. The method that we have presented, relies on
the different topologies and diagrams that can be derived at this
level. We have shown that relevant diagrams can be sorted in three,
and only three, categories. Thus, complemented with the SM quantum
numbers of the new BSM fields they provide a catalog for neutrino mass
models at the two-loop level. 

Finally, calculation of the corresponding neutrino mass matrices
involve at most four different types of two-loop integrals. We have
studied their behavior in a simple model, which serves as well to
illustrate how the results of the catalog can be used for two-loop
neutrino mass model building.
\section*{Acknowledgments}
I would like to thank my collaborators Audrey Degee, Luis Dorame and
Martin Hirsch. I acknowledge financial support from the FNRS agency
through a ``Charg\'e de Recherche'' contract.

\end{document}